\documentclass[conference]{IEEEtran}
\usepackage{cite}
\usepackage{amsmath,amssymb,amsfonts}
\usepackage{url}
\usepackage{algorithmic}
\usepackage{graphicx}
\usepackage{textcomp}
\usepackage{xcolor}
\usepackage{subcaption}
\usepackage[inline]{enumitem}
\usepackage{tasks}
\usepackage{balance}

\usepackage{mdframed}
\usepackage{newfloat} 
\usepackage{fourier}

\usepackage{caption}
\usepackage[bookmarks,bookmarksnumbered,
    pdfborder={0 0 0},
    linktocpage,
    colorlinks=true]{hyperref}

\usepackage{listings}

\def\BibTeX{{\rm B\kern-.05em{\sc i\kern-.025em b}\kern-.08em
    T\kern-.1667em\lower.7ex\hbox{E}\kern-.125emX}}

\begin{document}
\title{Model-based Design Tool for Cyber-physical Power Systems using SystemC-AMS}

\author{
\IEEEauthorblockN{Rahul Bhadani\IEEEauthorrefmark{1},
Satyaki Banik\IEEEauthorrefmark{2},
Hao Tu\IEEEauthorrefmark{2},
Srdjan Lukic\IEEEauthorrefmark{2i} and
Gabor Karsai\IEEEauthorrefmark{3}}
\IEEEauthorblockA{\IEEEauthorrefmark{1}Electrical \& Computer Engineering\\
The University of Alabama in Huntsville, Huntsville, AL, USA\\
Email: rahul.bhadani@uah.edu}
\IEEEauthorblockA{\IEEEauthorrefmark{2}
North Carolina State University, Raleigh, NC, USA\\
Email: sbanik@ncsu.edu, htu@ncsu.edu, smlukic@ncsu.edu}
\IEEEauthorblockA{\IEEEauthorrefmark{3}
Vanderbilt University, Nashville, TN, USA\\
Email: gabor.karsai@vanderbilt.edu}
}

\maketitle

\begin{abstract}
Cyber-physical power systems, such as grids, integrate computational and communication components with physical systems to introduce novel functions and improve resilience and fault tolerance. These systems employ computational components and real-time controllers to meet power demands. Microgrids, comprising interconnected components, energy resources within defined electrical boundaries, computational elements, and controllers, offer a solution for integrating renewable energy sources and ensuring resilience in electricity demand. Simulating these cyber-physical systems (CPS) is vital for grid design, as it facilitates the modeling and control of both continuous physical processes and discrete-time power converters and controllers. This paper presents a model-based design tool for simulating cyber-physical power systems, including microgrids, using SystemC-AMS. The adoption of SystemC-AMS enables physical modeling with both native components from the SystemC-AMS library and user-defined computational elements. We observe that SystemC-AMS can accurately produce the electromagnetic transient responses essential for analyzing grid stability. Additionally, we demonstrate the effectiveness of SystemC-AMS through use cases that simulate grid-following inverters. Comparing the SystemC-AMS implementation to one in Simulink reveals that SystemC-AMS offers a more rapid simulation. A design tool like this could support microgrid designers in making informed decisions about the selection of microgrid components prior to installation and deployment.

\end{abstract}

\section{Introduction}
\label{sec:intro}
Distributed and renewable energy resources have transformed the power grid to meet electricity demand. Modern power grids include various computational and communication components integrated with digital controllers, which renders them suitable for study as Cyber-physical Systems (CPS). A microgrid is a small, localized version of the power grid, created to utilize distributed and/or renewable energy resources to meet electricity demand when the main grid is compromised by situations such as grid failure or cyber-attacks. When connected to the main grid, if there is a surplus, the microgrid can supply power back to it, thus imposing additional requirements on grid operators to enhance stability and flexibility.

Given the logistical challenges and costs associated with installing a microgrid, direct installation without prior study through computer simulation may not yield optimal outcomes. Several software programs exist for microgrid simulation, offering a range of capabilities including economic dispatch, secondary control, primary control, and electromagnetic transient (EMT) study~\cite{alzahrani2017modeling}.

Economic dispatch simulations may not require detailed modeling of microgrid components; however, conducting an EMT simulation necessitates comprehensive modeling that may include both physical circuit representations and mathematical models. In this paper, we adopt a model-based engineering design approach to develop grid components. These components abstract physical models, such as electrical circuits representing transmission lines, resistors, and capacitors, as well as mathematical models, such as transfer functions representing low-pass filters, and subsystems like phase-locked loops. While physical models are inherently continuous, their behavior is significantly affected by the discrete nature of the controllers. Therefore, it requires careful coordination of simulation parameters to ensure a high-fidelity Cyber-physical System (CPS) simulation.

The main contribution of this paper is the provision of a tool that facilitates the creation of abstracted components for microgrid simulation within SystemC-AMS, an analog and mixed-signal extension of the widely used SystemC library. This paper explores the relationship between the physical processes, the timing properties of the simulation, and the controller employed. We illustrate the use of SystemC-AMS to develop a grid-following inverter for photovoltaic (PV) systems that can track reference power based on a given load. The model-based design tool proposed here assists microgrid designers in analyzing various configurations, potentially reducing installation costs and time by informing decisions about grid component specifications.
\section{Related Work}
\label{sec:related_work}
The use of model-based design for modeling and simulating Cyber-physical Power Systems  (CPPS) is a recent phenomenon attributed to the inclusion of communication and computational components to make grid operation resilient and fault-tolerant. Most of the work that has been done in power systems modeling doesn't take into account the simulation aspects of cyber-physical systems. Electrical Power System Modelling in Modelica~\cite{mattsson1998physical} is one of the earlier works that led to some Modelica packages such as Spot, ObjectStab, and PowerSystems Library~\cite{winkler2017electrical}, OpenIPSL~\cite{baudette2018openipsl}. OpenIPSL supports phasor time-domain simulation. Modelica is useful as it supports objected-oriented programming, and is a widely popular multi-domain modeling language for cyber-physical systems.  In~\cite{cui2020hybrid}, authors propose a symbolic-numerical hybrid model for the simulation of power systems. Their work has led to the creation of an open-source tool ANDES from writing models from block diagrams using predefined blocks that further generate Python code using the symbolic Python library. Another library written in C++ is DPSIM~\cite{mirz2019dpsim} which performs dynamic phasor-based simulation for power systems. However, DPSIM doesn't provide an ability to write custom components or controller models and simulation is limited to only components available in the library. In~\cite{haugdal2021open}, authors present DynPSSimPy which is able to simulate small to medium-sized grids using Python. DynPSSimPy is designed for reproducibility and expandability rather than speed of execution and accuracy.

When modeling power systems for simulation as a CPS, we need to consider a tight interaction between physical modeling and discrete-event simulation for controlling required signals to meet desired objectives. For CPPS, as we see increased usage of power electronic converters in grid design, we need to model EMTs that require conducting a simulation with very high accuracy and at a much smaller time-step to observe the necessary transients. Using Modelica~\cite{llerins2022modelling}, researchers have modeled power system simulation to produce EMT. With Modelica-like tools, the challenge remains in terms of extensibility, learning curve, and integration with other existing tools. Python-based tools suffer from the speed of execution and generally lack hardware-in-the-loop simulation support.

In addition to EMT simulation, a simulation tool should be able to provide the ability to create custom models for controllers as well as physical components. With SystemC and SystemC-AMS, we can not only model power systems and microgrid components but also, as it is written in C++, a vast number of C++ libraries can be used alongside SystemC code to provide added functionality. Further, we are looking to incorporate power electronic components at varying levels of fidelity -- from physical modeling using electrical circuit components to mathematical models using discrete transfer functions or state-space equations. As such, we find SystemC-AMS suitable for modeling power systems using physical components such as resistors, capacitors, and controlled current generators in addition to modeling using transfer functions and state-space equations. Besides, we can also model software and embedded system components with SystemC which is not available with any of the tools discussed earlier~\cite{hartmann2009modeling}.

SystemC~\cite{IEEE16662023} uses the discrete-event model of computation and utilizes delta-cycle to achieve deterministic simulation results. Delta cycles are non-time-consuming time-steps that consume zero time in simulation after a preceding event. SystemC-AMS, a supplemental library in C++, provides the ability to model analog and mixed-signal components and performs simulation using the SystemC simulation kernel.

The underlying motivation behind this work is to propose an alternative tool for modeling CPPS that is a low-code solution, easily integrable with other widely available software packages, and is free, fast and open-source.

\section{Modeling CPPS in SystemC-AMS}
To create components for microgrids in SystemC-AMS, we utilize a model-based design tool called COSIDE~\cite{coside}. COSIDE offers low-code or no-code drag-and-drop support for primitives (i.e., predefined basic components in SystemC-AMS) and any user-defined modules, enabling the visual construction of CPPS. COSIDE generates SystemC and SystemC-AMS C++ code skeletons for the TDF (Timed Data Flow) Model of Computation (MoC), while for the ELN (Electrical Linear Network) MoC, it generates read-only SystemC code. Users can further edit the TDF-generated code and incorporate their logic within the \texttt{processing} function of TDF modules. ELN primitives can be selected and placed using a library in a schematic editor to build a subsystem or a complete system, which can then be reused in other schematic editors to create more complex systems.

To model a CPPS, it is essential to identify electrical components, any controllers such as a phase-locked loop, environmental characteristics, and relevant parameters. Additionally, determining the maximum latency that components in the system can tolerate, as well as the simulation step size that accommodates this maximum latency, is crucial for ensuring the accurate propagation of results across channels between modules.

\subsection{Physical-modeling through ELN MoC}
\label{sec:ELN_MoC}
\begin{figure*}[htpb]
\centering
\includegraphics[width=0.85\linewidth, trim={4.0cm 8.0cm 4.0cm 4.0cm},clip]{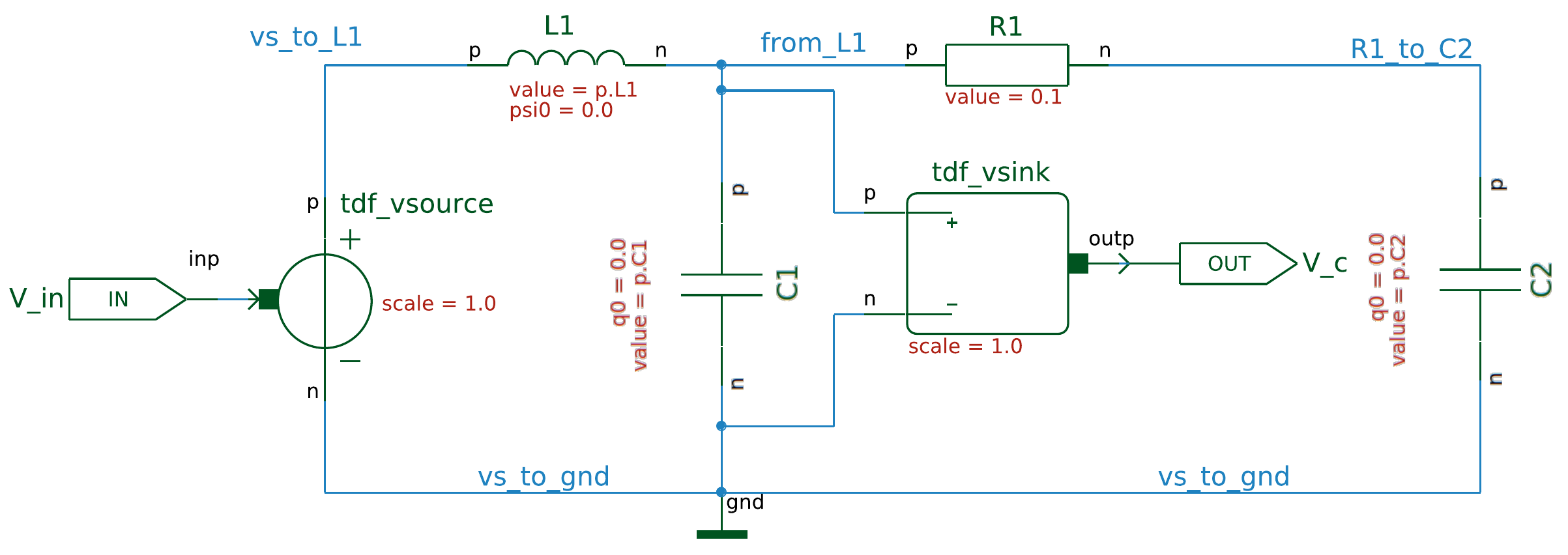}
\caption{A circuit schematic illustrating a physical model using ELN MoC. The voltage source receives a discrete-time signal as an input from a TDF block. We measure the output as the voltage across the capacitor $C1$ using another ELN primitive voltage sink which acts as a voltmeter. Input and output ports are shown as elongated pentagonal boxes, labeled \texttt{IN} and \texttt{OUT}. The overall circuit with input/output ports can be abstracted as shown in Figure~\ref{fig:EMCircuit_Block}.}
\label{fig:ELN_Circuit}
\end{figure*}

\begin{figure}[htpb]
\centering
\includegraphics[width=0.99\linewidth, trim={0cm 0.0cm 0cm 0.0cm},clip]{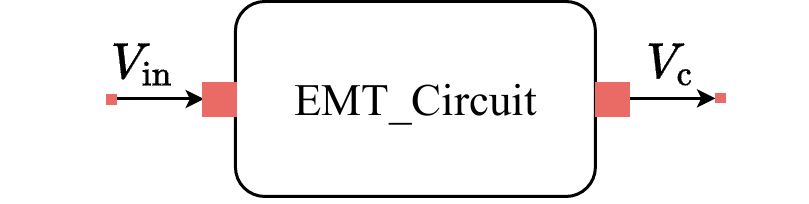}
\caption{An abstraction of Figure~\ref{fig:ELN_Circuit} that can be reused as a subsystem.}
\label{fig:EMCircuit_Block}
\end{figure}

SystemC-AMS enables the modeling of physical current and voltage sources with its current and voltage source primitive modules, which epitomize continuous-time entities. Furthermore, SystemC-AMS allows for the creation of controlled current and voltage sources, which are governed by inputs from discrete-time TDF modules, as discussed in Section~\ref{sec:TDF_MoC}. An illustrative example of a circuit representing a physical model can be found in Figure~\ref{fig:ELN_Circuit}. Components can be added to the circuit by dragging and dropping them into the COSIDE editor's workspace. The editor then generates a SystemC module, with a class declaration exemplified in Listing~\ref{code:ELN_Circuit}. The \texttt{architecture} function details the interconnections between the circuit's components. The remaining code is omitted for brevity but will be available as open-source with the paper's accepted version. The designed circuit is abstracted as a library component, which can be visualized in Figure~\ref{fig:EMCircuit_Block}, making it available for reuse as a subsystem.

\begin{mdframed}[linecolor=black, topline=false, bottomline=false,
  leftline=false, rightline=false, backgroundcolor=white]
\begin{lstlisting}
SC_MODULE(EMT_Circuit)
{
     /// ports
     sca_tdf::sca_out<double> V_c;
     sca_tdf::sca_in<double> V_in;
    /// parameters
    struct params
    {
        double SIM_STEP; /** Simulation Step */
        unsigned int T_RES; /** Time Resolution */
        double L1; /** Inductance in Henry */
        double C1; /** Capacitance in Farad */
        double R1; /** Resistance in Ohm */
        double C2; /** Capacitance in Farad */
        params()
        {
            SIM_STEP = 0.000017067;
            T_RES = sc_core::SC_MS;
            L1 = 0.0022;
            C1 = 5.1200e-06;
            R1 = 0.1;
            C2 = 1.0240e-05;
        }
    } p;
    void architecture();
    EMT_Circuit(sc_core::sc_module_name, 
    	const params& pa = params());
    virtual ~EMT_Circuit();
    struct components;
      components* c;
      components* operator -> () { return c; }
};

\end{lstlisting}
\captionof{lstlisting}{C++ Declaration of SystemC module corresponding to the circuit shown in Figure~\ref{fig:ELN_Circuit}. C++ structure \texttt{components} specifies components of the circuit, and the function \texttt{architecture} specifies how the terminals of each component are connected with each other.}
\label{code:ELN_Circuit}
\end{mdframed}

\subsection{Discrete-time models through TDF MoC}
\label{sec:TDF_MoC}
TDF MoC is adept at modeling discrete-time systems and facilitating their simulations without the need for the dynamic scheduling required by SystemC's discrete event kernel. When TDF modules are interconnected, they form a TDF cluster with a static schedule, which defines the execution sequence, regulates the number of samples to be read from or written to ports, and specifies delays at those ports. These port delays are critical in resolving algebraic loops within feedback systems and can emulate sensing delays. Like the component shown in Figure~\ref{fig:EMCircuit_Block}, these TDF modules can be abstracted and reused to construct hierarchically complex systems, with input and output ports that can be connected to other subsystems and parameterized according to user-defined specifications at the time of design.
\section{Use Cases}
\label{eq:usecases}
In this section, we use SystemC-AMS and the model-based design framework previously discussed to several use cases. We use the circuit detailed in Figure~\ref{fig:ELN_Circuit} to demonstrate EMT analysis in an open-loop system. Additionally, we simulate a grid-following inverter, which is a pivotal component of a microgrid. To validate the correctness of our approach, we compare the simulation results of the grid-following inverter with an equivalent implementation in Simulink.

\subsection{Electromagnetic Transients in the Simulation for Open-loop System}
The electrical circuit shown in Figure~\ref{fig:ELN_Circuit} is an equivalent electromagnetic transient model of a wye-connected, solidly-grounded three-phase transformer with two capacitor banks~\cite{watson2003power, zhao2019electromagnetic, su2012electromagnetic}. 
Referencing the circuit in Figure~\ref{fig:ELN_Circuit}, connecting the two capacitors in parallel leads to a rapid transfer of charge between them. The time constant associated with this transfer is 
\begin{equation}
    \begin{split}
        \tau = R\bigg( \frac{C_1C_2}{C_1 + C_2}\bigg) = 3.35 \times 10^{-7} \textrm{~sec}
    \end{split}
\end{equation}
if $C_1 = 5.12 \mu F$, and $C_2 = 10.24 \mu F$. The parallel capacitance, when combined with the inductance, results in a natural response at a frequency of  
\begin{equation}
    \begin{split}
        \omega_0 & = \cfrac{1}{\sqrt{L(C_1 + C_2)}} = 5440 \textrm{rad/s} \quad \Rightarrow T  = \cfrac{2\pi}{\omega_0} = 1.15 ~\textrm{ms}
    \end{split}
\end{equation}
for the inductance value $L = 2.2~\textrm{mH}$. 

\begin{figure}[htpb]
\centering
\includegraphics[width=0.99\linewidth, trim={0cm 0.0cm 0cm 0.0cm},clip]{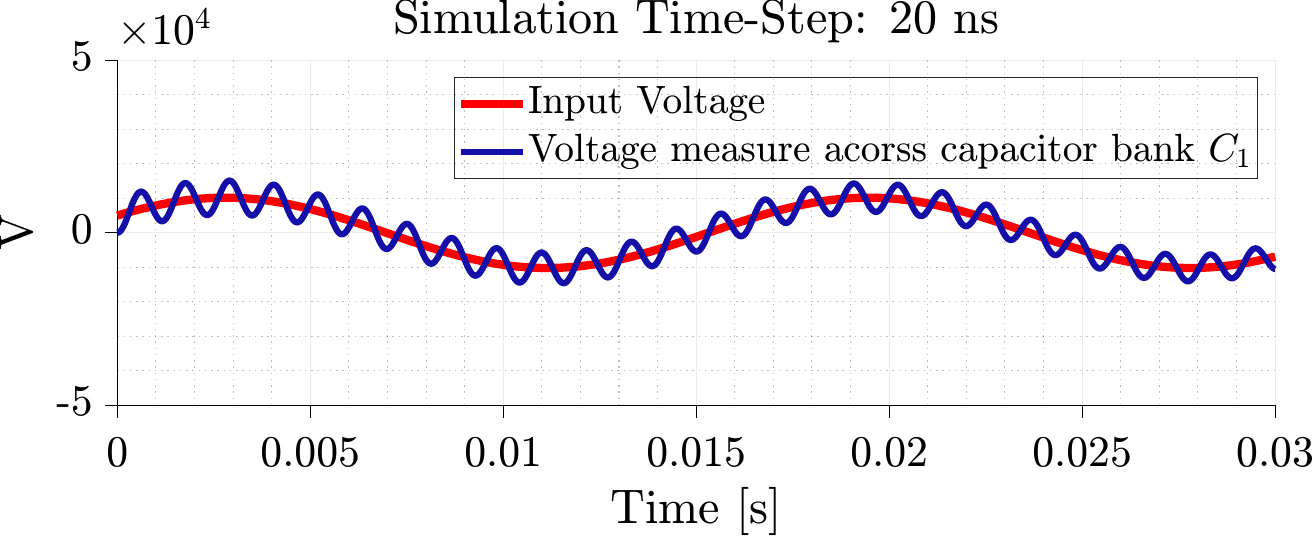}
\caption{EMT phenomenon observed in the voltage measured across Capacitor bank $C_1$ from the circuit in Figure~\ref{fig:ELN_Circuit} when simulated in SystemC-AMS with the time-step of $20~\textrm{ns}$. The transient eventually disappears at around $0.35~\textrm{s}$ (not shown in the figure), and the voltage across the capacity $C_1$ stabilizes.}
\label{fig:EMT_Simulation_20ns}
\end{figure}

\begin{figure}[htpb]
\centering
\includegraphics[width=0.99\linewidth, trim={0cm 0.0cm 0cm 0.0cm},clip]{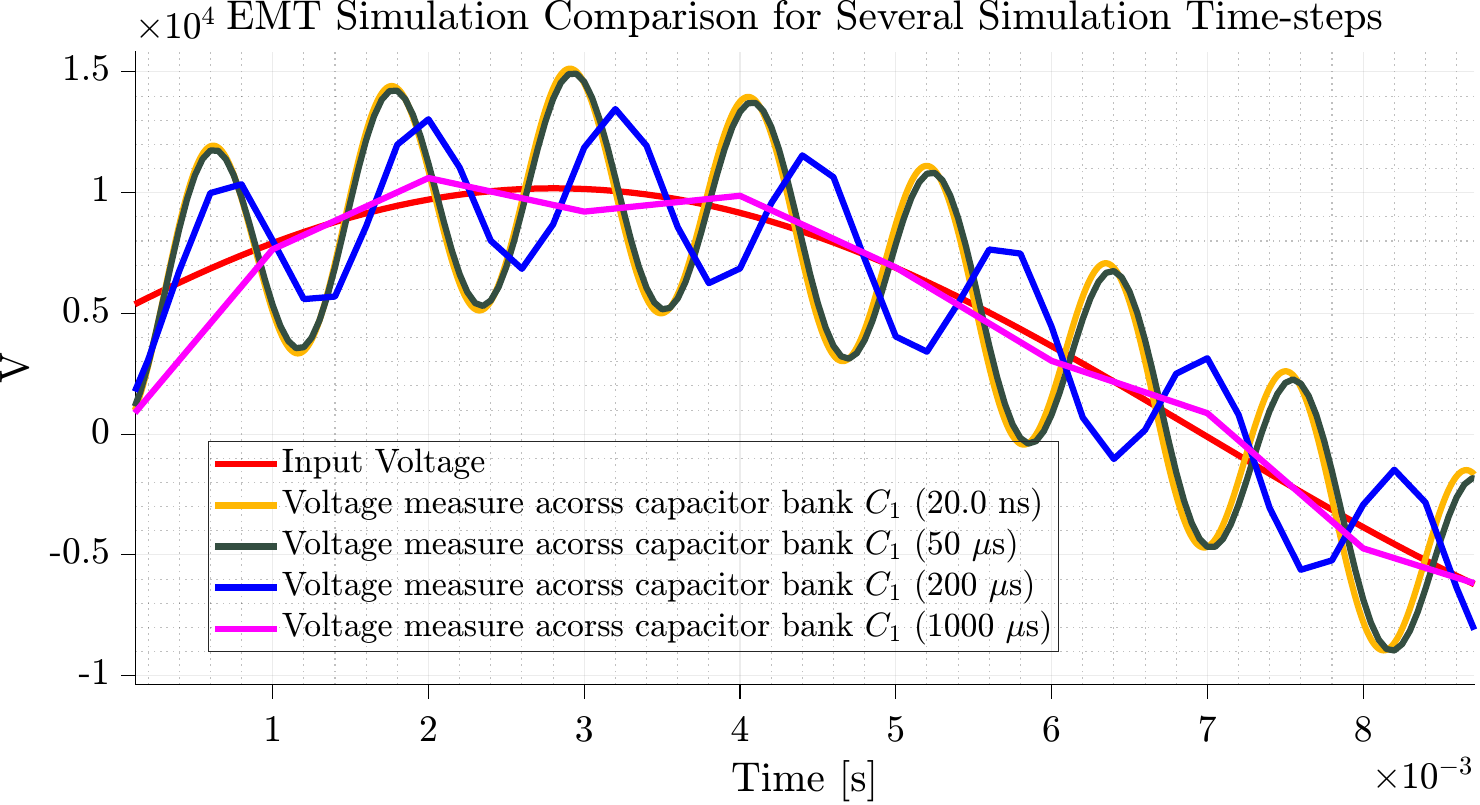}
\caption{EMT phenomenon observed in the voltage measured across Capacitor bank $C_1$from the circuit ~\ref{fig:ELN_Circuit} when simulated in SystemC-AMS with several different time-step values. Simulation with a large time step fails to produce an EMT phenomenon while a simulation with a time step smaller than the time period of the natural frequency exhibits an EMT phenomenon.}
\label{fig:EMT_Simulation}
\end{figure}

In discrete-time-step simulators used for EMT response, the time step is typically much smaller than the natural response period of the circuit to accurately capture the circuit's transient behavior. A commonly employed rule of thumb is to select a time step that is at least an order of magnitude smaller than the smallest time constant in the system~\cite{liu2022using, wu2023fractional}. To initiate a transient, we select a simulation time step of \(0.05~\textrm{ms}\) or \(50~\mu\textrm{s}\). Utilizing an exceedingly small time step on the order of nanoseconds leads to slower simulation speeds but yields the most precise results, as depicted in Figure~\ref{fig:EMT_Simulation_20ns}. Conducting the simulation with a substantially smaller time step, while focusing on a narrow region of the input/output signals, reveals that any simulation time step exceeding the system's time constant will not accurately capture the EMT, as shown in Figure~\ref{fig:EMT_Simulation}. The figure also suggests that simulating with a time step in the nanosecond range is unnecessary for observing EMT, as further reduction in the time step yields no additional benefit. However, it is important to point out that simulations with a larger time step, such as \(50~\mu \textrm{s}\), show a gradual phase shift as the simulation progresses. Steady state is reached at around \(0.35~\textrm{s}\), which is not represented in the figure. You can observe this behavior in the voltage measurement across capacitor bank \(C_1\) illustrated in Figure~\ref{fig:EMT_Simulation}. We find that the phase-shift is not of concern for a feedback-loop-based system where a controller is designed to correct the tracking of a reference signal and the main purpose of the simulation centers around tackling high-frequency transients and a slight phase shift doesn't obstruct such a study.

\subsection{PV-based Grid-following Inverter Design for Microgrid}
Grid-following (GFL) control is commonly employed in grid-connected inverters, enabling the inverter to function akin to a current source. The principal aim of a GFL inverter is to synchronize with the grid's frequency and to operate as a regulated current source at a designated power output. It is designed to deliver the necessary quantities of active and reactive power to the main grid. GFL inverters are capable of maintaining nearly constant output currents or power levels despite load variations. This fine-tuning of active and reactive power is achieved by monitoring the grid voltage, utilizing a Phase-Locked Loop (PLL)~\cite{dong2014analysis, kamal2018three}, and a current control loop, which allows rapid adjustment of the GFL inverter's output current.

The conventional approach for implementing a linear controller in a three-phase system involves a PI (Proportional-Integral) controller operating in a dq-synchronous reference frame. This setup includes two separate control loops that manage the direct and quadrature components. However, most commercially available dynamic stability simulation tools model grid-following inverters as adjustable current sources, disregarding the inner control loops \cite{nrel2014PSCADmodelforPV, PSSE}. This research treats GFL inverters as photovoltaic (PV) units, depicting the inverter side with an adjustable three-phase current source combined with a parasitic resistance and a high-value snubber capacitance. This configuration is designed to absorb and dissipate high-frequency oscillations, minimize overshooting, and enhance the inverter’s overall transient response \cite{xue2022siemens,du2021GFLGFM}.

\begin{figure}[htpb]
\centering
\includegraphics[width=0.99\linewidth, trim={0cm 0.0cm 0cm 0.0cm},clip]{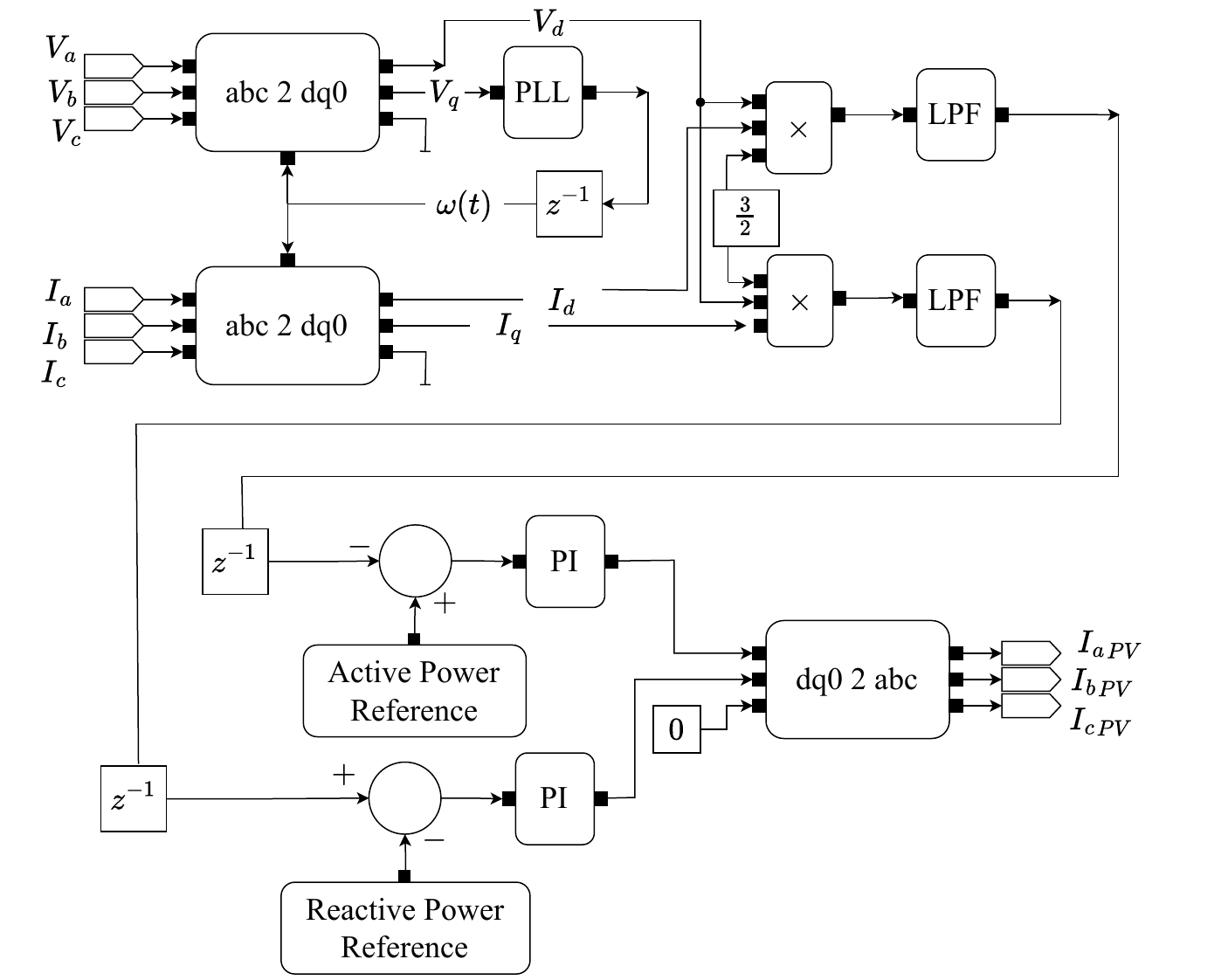}
\caption{Simplified GFL inverter without inner current loops. To break the algebraic loop that arises due to feedback, we use a delay unit $z^{-1}$ which introduces the port delay by one sample at the respective output port. LPF stands for low-pass filter. The GFL inverter model is abstracted as PV-GFL and works with the main grid as shown in Figure~\ref{fig:Microgrid_architecture}. \texttt{abc 2 dq0} block is a TDF module that converts a time-varying three-phase signal to a dq0 reference frame (where signals appear to have a constant phase). \texttt{dq0 2 abc} does the opposite job. A PI block is a discrete-time controller implemented using TDF MoC with a sampling frequency of $1000~\textrm{Hz}$. We can specify active power reference and reactive power reference for the simulation as time-domain functions which are implemented using TDF MoC.}
\label{fig:PVGFL_LPF}
\end{figure}

In the $dq$ reference frame, to control, aligning the $d$-axis with the space phasor of the plant model is crucial \cite{IravaniBook}. The PLL aims to lock the GFL into the grid’s frequency or phase through a feedback implementation that nullifies the $q$-axis component of the inverter output voltage, $V_{sq}$. Consequently, the $d$-axis component of the inverter output voltage, $V_{sd}$, equals the RMS output voltage, $\hat{V}$.
\begin{equation}
V_{sd} = \hat{V}, V_{sq} = 0
\end{equation}
The GFL’s control objective is to manage the real power, $P_s$, and the reactive power, $Q_s$, injected into the grid from the inverter.
\begin{equation}
\begin{split}
P_s(t) & = \frac{3}{2}[V_{sd}(t) i_d(t) + V_{sq}(t) i_q(t)]\\
Q_s(t) & = \frac{3}{2}[-V_{sd}(t) i_q(t) + V_{sq}(t) i_d(t)]
\end{split}
\label{eq:power_eq1}
\end{equation}
Given that $V_{sq}=0$,
\begin{equation}
\begin{split}
P_s(t) & = \frac{3}{2} V_{sd}(t) i_d(t), \quad
Q_s(t) = -\frac{3}{2} V_{sd}(t) i_q(t)
\end{split}
\label{eq:power_eq2}
\end{equation}
From these equations, we derive separate current references in the $dq$ domain,
\begin{equation}
\begin{split}
i_{d,ref}(t) = \frac{2}{3 V_{sd}} P_{s,ref}(t) , \quad
i_{q,ref}(t) = -\frac{2}{3 V_{sd}} Q_{s,ref}(t)
\end{split}
\label{eq:Iref}
\end{equation}
where $ref$ notation is used for reference signals.

In this paper, we discuss a simplified GFL inverter without inner current loops. Commercial dynamic simulation software typically models GFL inverters as controllable current sources, excluding the inner current loops. In the absence of the inner current loop, the control scheme mainly comprises the outer power loop, which produces the current references for the controllable current sources. A simplified block diagram of the GFL inverter, devoid of inner current loops, is depicted in Fig.~\ref{fig:PVGFL_LPF}.

The active and reactive power, denoted by $P$ and $Q$, are measured in the $dq$-reference frame as per Eq.~\eqref{eq:power_eq1} and subsequently filtered using a discrete low-pass filter to eliminate high-frequency elements in power measurement. The bandwidth of this low-pass filter can be further utilized to incorporate an inertial response from the inverter \cite{Poolla2019FFR}. The $d$ and $q$ axes are uncoupled, thereby enabling independent control over $P$ and $Q$. However, due to the EMT phenomenon, we see some effect of step-change in active power on the reactive power.  The measured power is then subtracted from the references, $P_\textrm{ref}$ and $Q_\textrm{ref}$, and passed through a PI controller that monitors the deviation of active and reactive power from the set reference to stabilize the instantaneous active and reactive power. 

\begin{figure}[htpb]
\centering
\includegraphics[width=1.0\linewidth, trim={0cm 0.0cm 0cm 0.0cm},clip]{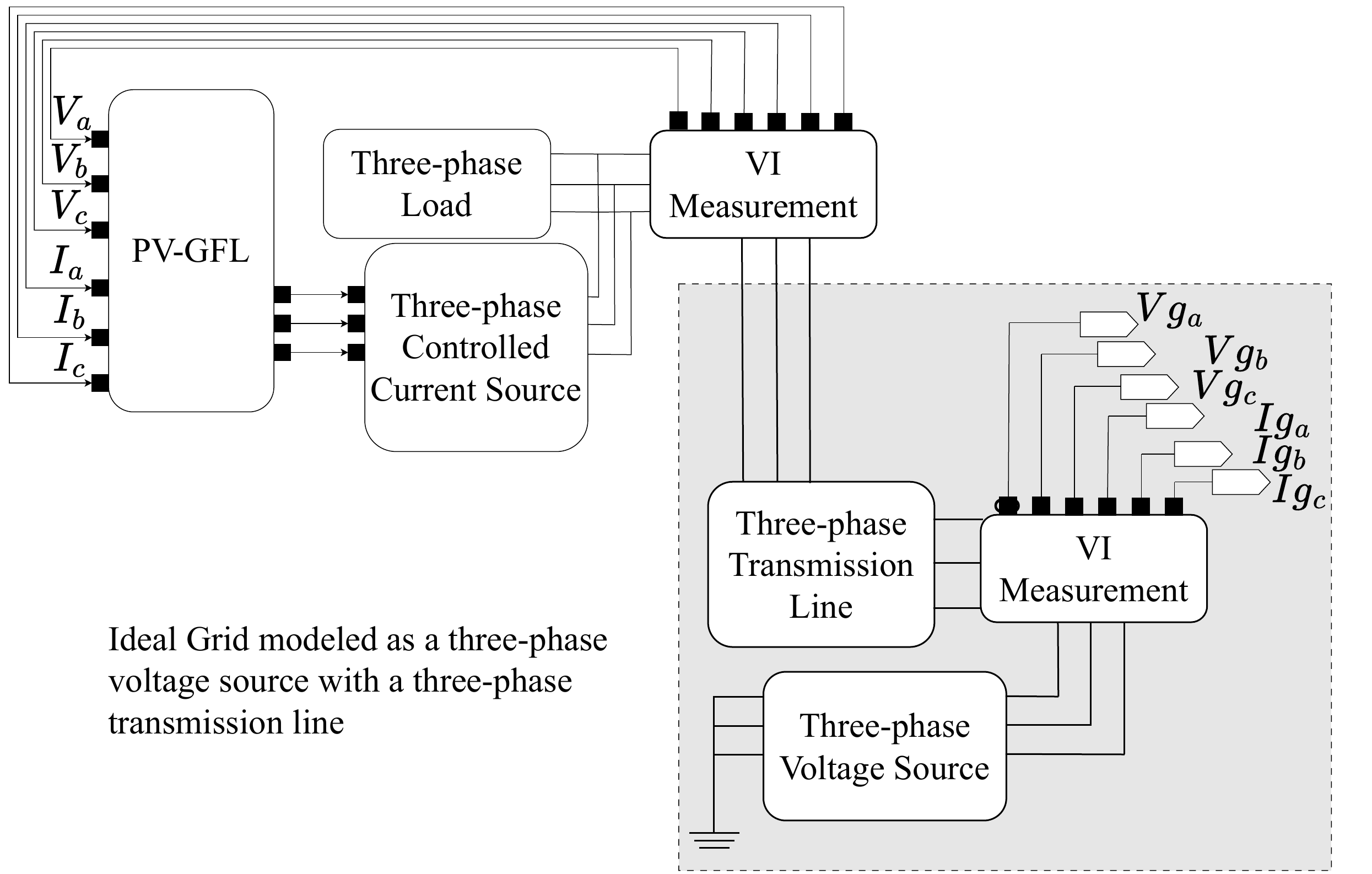}
\caption{A PV-GFL inverter acting as a microgrid that works with the main grid. The main grid is modeled as a three-phase voltage source connected through the transmission line. Three-phase controlled current source is a group of three controlled current source elements provided by the SystemC-AMS library that act as a current source based on the input value. VI measurement block consists of a voltmeter and an ammeter (called voltage sink and current sink in SystemC-AMS) that measure voltage and current respectively. The overall implementation also provides three-phase grid voltage $V_{g_a, g_b, g_c}$ and current $I_{g_a, g_b, g_c}$ as output that can be logged during the simulation.}
\label{fig:Microgrid_architecture}
\end{figure}

We can use the PV-GFL design for constructing a microgrid and its interaction with the main grid. The use of PV-GFL for operating with the main grid is depicted in  Figure~\ref{fig:Microgrid_architecture}. PV-GFL model receives three-phase voltage $V_{a, b, c}$ and three-phase current $I_{a, b, c}$ as inputs and outputs regulated current $I_{aPV, bPV, cPV}$ as output. Figure~\ref{fig:Microgrid_architecture} has an ideal grid modeled as a three-phase voltage source connected through a transmission line. The transmission is modeled as a lossy transmission using resistance in series with inductance. The overall model comprises an algebraic loop which can be broken by introducing a delay unit $z^{-1}$ (where is $z$ variable is from z-transform) in the GFL inverter model as shown in Figure~\ref{fig:PVGFL_LPF}. We measure the instantaneous active and reactive power $P$, and $Q$ using three-phase voltages $V_{a, b, c}$, and currents $I_{a, b, c}$ using the formula from Equation~\eqref{eq:active_reactivePQ} as follows:
\begin{equation}
\label{eq:active_reactivePQ}
    \begin{split}
        P & = V_aI_a + V_b I_b + V_cI_c\\
        Q & = \cfrac{1}{\sqrt{3}}\bigg( ( V_a - V_b)I_c + (V_b - V_c)I_a + (V_c  - V_a)I_b \bigg)
    \end{split}
\end{equation}

\begin{figure}[htpb]
\captionsetup[subfigure]{aboveskip=0pt,belowskip=1pt}
\centering
\begin{subfigure}{1.0\linewidth}
\centering
\includegraphics[width=1.0\linewidth, trim={0.0cm 0.0cm 0.0cm 0.0cm},clip]{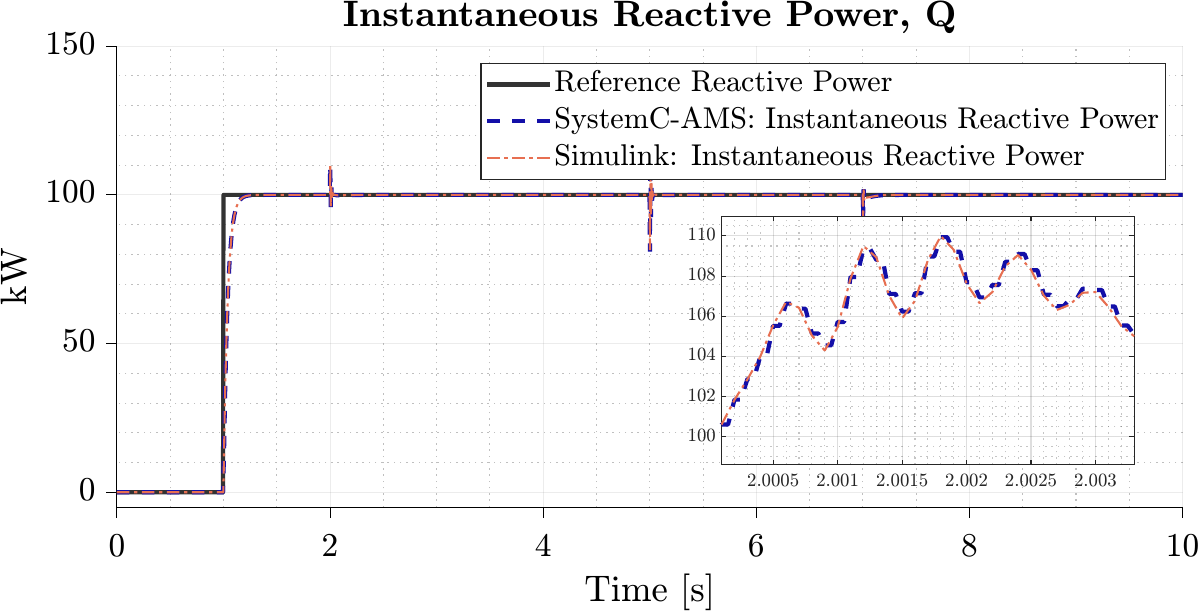}
\caption{}
\end{subfigure}
\begin{subfigure}{1.0\linewidth}
\centering
\includegraphics[width=1.0\linewidth, trim={0.0cm 0.0cm 0.0cm 0.0cm},clip]{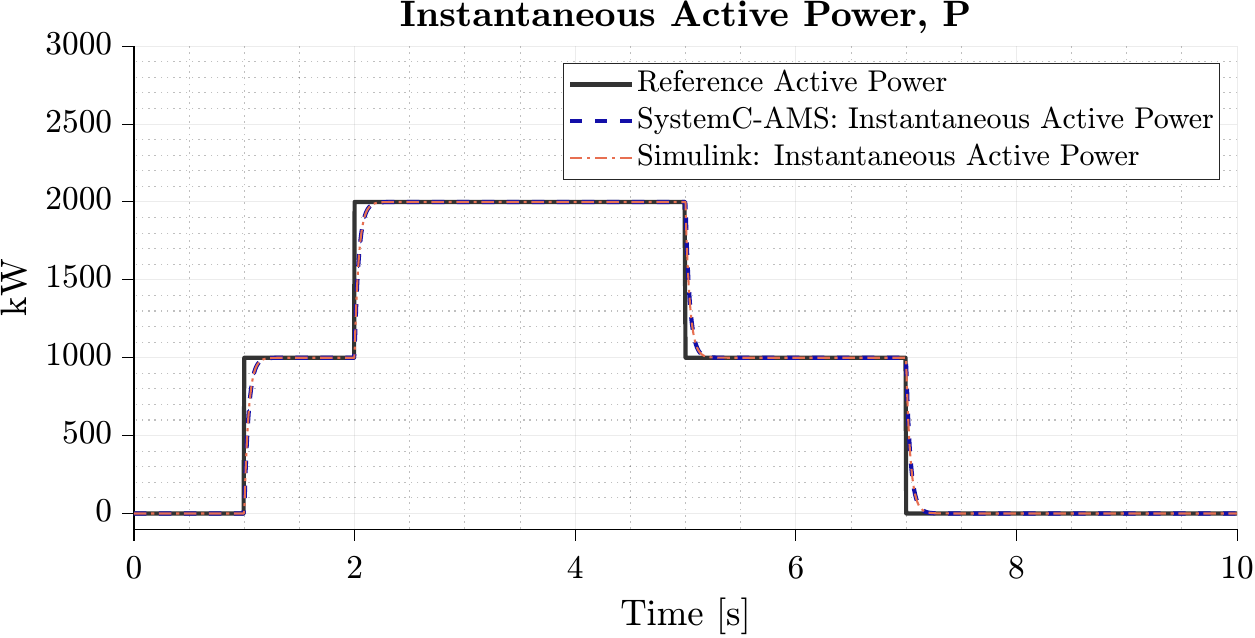}
\caption{}
\end{subfigure}
\caption{\textbf{\textit{Simplified GFL inverter without inner current loops: step response.}} \textbf{Top: }Instantaneous reactive power plot with SystemC-AMS simulation of GFL inverter-based microgrid. We zoom in around the first transients as you can see in the inset figure.  We also plot the reference reactive power. \textbf{Bottom: }Instantaneous active power plot in SystemC-AMS simulation. Notice the electromagnetic transients in reactive power as there is a step change in the active power -- thereby demonstrating the capability to simulate EMT. Traces of signals captured in Simulink implementation also closely match with what we see in SystemC-AMS implementation.}
\label{fig:power_plot_lpf}
\end{figure}

To conduct the simulation in SystemC-AMS, we use a simulation time-step of $50~\mu\textrm{s}$ and run the simulation for $10~\textrm{s}$. The three-phase voltage source has the root-mean-square phase-to-phase voltage of $480$~V operating at $60$ Hz. The transmission line~\cite{chew2020lectures} is modeled as a series resistance of $0.01~\Omega$, a series inductance of $0.0001~H$, a shunt resistance of $0.15~\Omega$, and shunt capacitance of $80~\mu F$. Three-phase load is a pure resistive load of $1000~\Omega$. The low-pass filter (LPF block in Figure~\ref{fig:PVGFL_LPF}) uses the z-domain transfer function $\tfrac{0.0609}{z - 0.9391}$ at the sampling rate of $1000~\textrm{Hz}$. Phase-locked-loop (PLL block in Figure~\ref{fig:PVGFL_LPF}) establishes a relationship between grid voltage and frequency. A GFL inverter uses PLL to keep the inverter in synchronization with the main grid. The measured angle is used to control the current. We study the step response by providing several step-change inputs that act as references for reactive and active power. We observe electromagnetic transients in the reactive power in response to the change in the active power as shown in Figure~\ref{fig:power_plot_lpf}. Finally, we also compared the implementation of the GFL inverter design in SystemC-AMS with one in Simulink and our implementation demonstrated that SystemC-AMS provides three times faster simulation compared to Simulink-based simulation.

The RMS error of the instantaneous active power compared to the reference active power for SystemC-AMS simulation is $85.78328~\textrm{kW}$, and $85.87311~\textrm{kW}$ for Simulink simulation which is roughly within $1\%$ the initial step amplitude of $1000~\textrm{kW}$. The reactive power RMS error was found to be $4.35805~\textrm{kVar}$, and $4.36262~\textrm{kVar}$ for SystemC-AMS simulation and Simulink simulation respectively. In addition, we also assess the RMS value for reactive power between two simulation methods around the time when the first transient occurs. The first transient takes approximately 0.2 seconds to stabilize. RMS error between Simulink and SystemC-AMS simulink, while transient lasted, came out to be $0.04996~\textrm{kVar}$.

\begin{figure}[htpb]
\centering
\includegraphics[width=1.0\linewidth, trim={0cm 0.0cm 0cm 0.0cm},clip]{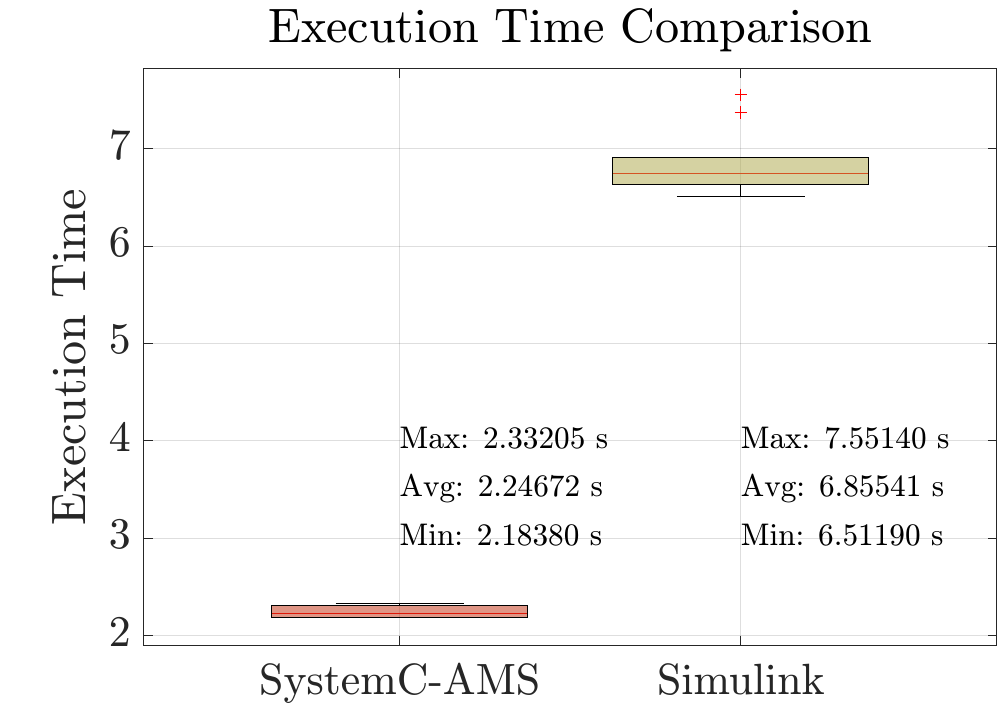}
\caption{Execution time comparison of simulation conducted in SystemC-AMS and Simulink We observe that the simulation conducted in SystemC-AMS is approximately 3x faster than one done in Simulink.}
\label{fig:boxplot}
\end{figure}

In terms of run-time performance, our calculation shows that the simulation done in SystemC-AMS was approximately three times faster than the one done in Simulink. Figure~\ref{fig:boxplot} illustrates the execution performance using a boxplot diagram.
\section{Conclusion and Future Work}
In this paper, we have presented a model-based design tool for simulating microgrid components using SystemC-AMS, constructing a DC microgrid, and a microgrid design using GFL inverters. The simulation study of a GFL inverter with an ideal main grid has shown that SystemC-AMS can facilitate rapid simulation, display the EMT phenomenon, and model microgrid components with high accuracy. This enables SystemC-AMS to act as a digital twin for microgrids, aiding in the development of prototypes and the refinement of control algorithms. Our future endeavors will focus on expanding the range of grid components in SystemC-AMS to facilitate large-scale microgrid studies and demonstrating real-time simulation capabilities in combination with hardware components to control grid signals under diverse conditions.

\section{Acknowledgements}
This material is based upon work supported by the Advanced Research Projects Agency-Energy (ARPA-E), U.S. Department
of Energy, under Award Number DE-AR0001580. The views expressed herein do not necessarily represent the views of the U.S. Department of Energy or the United States Government.

\bibliographystyle{IEEEtran}
\bibliography{references}

\end{document}